\documentclass[aps,prl,showpacs,reprint]{revtex4-1}
\usepackage{amsmath}
\usepackage{amssymb}
\usepackage{dcolumn}
\usepackage{graphicx}
\usepackage[colorlinks=true,linkcolor=blue,citecolor=blue]{hyperref}

\newcommand{\innC}{C_{0}}
\newcommand{\innS}{S_{0}}
\newcommand{\outC}{C^{\star}}
\newcommand{\outS}{S^{\star}}
\newcommand{\onaxisJ}{J_{0}}
\newcommand{\rbound}{r_{\text{B}}}
\newcommand{\plasmaI}{I_{\text{p}}}
\newcommand{\normpsi}{\psi_{\text{N}}}
\newcommand{\odd}[2]{#1_{#2}^{(s)}}
\newcommand{\even}[2]{#1_{#2}^{(c)}}
\newcommand{\shortsection}[1]{\textit{#1}.---}

\begin{document}

\title{Conditions for up-down asymmetry in the core of tokamak equilibria}

\date{\today}

\author{Paulo Rodrigues}
\email{par@ipfn.ist.utl.pt}
\author{Nuno F.~Loureiro}
\affiliation{
Associa\c{c}\~{a}o Euratom--IST,
Instituto de Plasmas e Fus\~{a}o Nuclear,
Instituto Superior T\'{e}cnico,
Universidade de Lisboa,
1049-001 Lisboa, Portugal.}

\author{Justin Ball}
\author{Felix I.~Parra}
\affiliation{
Plasma Science and Fusion Center, Massachusetts Institute of Technology,
Cambridge, Massachusetts, USA.}

\begin{abstract}

A local magnetic equilibrium solution is sought around the magnetic axis in
order to identify the key parameters defining the magnetic-surface's up-down
asymmetry in the core of tokamak plasmas. The asymmetry is found to be
determined essentially by the ratio of the toroidal current density flowing on
axis to the fraction of the external field's odd perturbation that manages to
propagate from the plasma boundary into the core. The predictions are tested and
illustrated first with an analytical Solovev equilibrium and then using
experimentally relevant numerical equilibria. Hollow current-density
distributions, and hence reverse magnetic shear, are seen to be crucial to bring
into the core asymmetry values that are usually found only near the plasma edge.

\end{abstract}


\maketitle

\shortsection{Introduction}%
Plasma turbulence is known to degrade particle and energy confinement in fusion
devices, with serious consequences on their performance~\cite{doyle.2007}.
However, turbulent transport can be reduced, or even suppressed, in toroidally
rotating plasmas by velocity gradients~\cite{doyle.2007, burrell.1997,
biglari.1990, waltz.1995, highcock.2010, barnes.2011}. In the absence of
external momentum sources, spontaneous (or intrinsic) plasma rotation may arise
due to the momentum flux induced by symmetry breaking along magnetic field
lines~\cite{peeters.2005, parra.2011, peeters.2011, camenen.2009a,
camenen.2009}. One such symmetry breaking mechanism is yielded by up-down
asymmetric equilibria. Unfortunately, early assessments have found that the
asymmetry due to the externally shaped plasma boundary largely fails to
propagate in to the core~\cite{camenen.2009a, camenen.2009, peeters.2011}. The
benefits of the induced momentum flux thus appear to be restricted to the outer
part of the plasma, with limited success in reducing turbulent transport levels.

In this Letter, a local analytic equilibrium is developed near the magnetic axis
in order to understand the factors on which the shape of the magnetic surfaces
depends and how the asymmetry may be enhanced. We find that reverse magnetic
shear configurations can significantly increase the asymmetry on axis and are
therefore expected to extend the asymmetry-induced momentum flux deep into the
core of tokamak plasmas, thereby improving its confinement properties.

\shortsection{Local on-axis equilibria}%
The distribution of the poloidal-field flux $\psi$, normalized to an arbitrary
constant $\normpsi$, in the poloidal section of axisymmetric devices is given by
the Grad-Shafranov (GS) equation~\cite{freidberg.1987,wesson.1997}
\begin{equation}
    - \frac{1}{r} \frac{\partial}{\partial r}
        \frac{r}{R} \frac{\partial \psi}{\partial r}
    - \frac{1}{r^{2}} \frac{\partial}{\partial \theta} \frac{1}{R}
        \frac{\partial \psi}{\partial \theta} =
        R \, \dot{p}( \psi ) +\frac{\dot{Y}( \psi )}{R},
\label{eq:grad.shafranov}
\end{equation}
if centrifugal effects are neglected. The static pressure $p(\psi)$ and the
squared poloidal current $Y(\psi)$ are normalized to $\mu_{0}^{-1} \normpsi^{2}
\big/ (a R_{0})^{2}$ and $8 \pi^{2} \normpsi^{2} \big/ (\mu_{0} a)^{2}$
respectively, the dots denote flux derivatives $d/d\psi$, and $\bigl( r, \theta,
\phi \bigr)$ are right-handed coordinates with the origin displaced by $R_{0}$
from the tokamak's symmetry axis: The distance $r$ to the origin is normalized
to $a$, $\theta$ is a poloidal angle measured from the midplane's high-field
side, $\phi$ is the toroidal angle, and $\varepsilon = a / R_{0}$ is the inverse
aspect ratio, with $a$ and $R_{0}$ the tokamak's minor and major radii. The
distance $R$ to the symmetry axis and the height $Z$ above the midplane, both
normalized to $R_{0}$, are
\begin{equation}
    R = 1 - \varepsilon r \cos \theta
        \quad \text{and} \quad Z = \varepsilon r \sin \theta.
\label{eq:RZ.transformation}
\end{equation}

We seek a solution to Eq.~\eqref{eq:grad.shafranov} in the
form~\cite{wesson.1997,rodrigues.2004,rodrigues.2009}
\begin{subequations}
\begin{gather}
    \psi( r, \theta ) = \psi_{0}(r) +
        \sum_{n = 1}^{\infty} \frac{\varepsilon^{n}}{n!} \psi_{n}(r,\theta), \\
    \psi_{n}( r, \theta ) = \sum_{k = 0}^{n}
            \even{\psi}{nk}(r) \cos k \theta + \odd{\psi}{nk}(r) \sin k \theta.
\end{gather}
\label{eq:ansatz}%
\end{subequations}
This \textit{ansatz} is valid for any $r$, $\theta$, and $\varepsilon$ as long
as both series converge, so $\psi_{n}(r,\theta) \sim 1$ is not formally
required. Replacing Eqs.~\eqref{eq:ansatz} in Eq.~\eqref{eq:grad.shafranov} and
collecting terms with equal powers of $\varepsilon$, one finds an equation for
$\psi_{0}$,
\begin{equation}
    r^{2} \psi_{0}'' + r \psi_{0}' =
        - r^{2} \Bigl[ \dot{p}_{0}(r) + \dot{Y}_{0}(r) \Bigr],
\label{eq:0th.ode}
\end{equation}
in which $p_{0}(r) = p \bigl[ \psi_{0}(r) \bigr]$, $Y_{0}(r) = Y \bigl[
\psi_{0}(r) \bigr]$, and the primes denote radial derivatives. Likewise, a
sequence of linear inhomogeneous equations,
\begin{equation}
    r^{2} \psi_{nk}'' + r \psi_{nk}' +
        \bigl[ r^{2} \sigma(r) - k^{2} \bigr] \psi_{nk} = b_{nk},
\label{eq:nkth.ode}
\end{equation}
is found for all other harmonics $\psi_{nk}$ (any of $\even{\psi}{nk}$ or
$\odd{\psi}{nk}$), with $\sigma(r) = \ddot{p}_{0}(r) + \ddot{Y}_{0}(r)$. Each
$b_{nk}(r)$ is the $k^{\text{th}}$ harmonic [either $\even{b}{nk}(r)$ or
$\odd{b}{nk}(r)$] of the $n^{\text{th}}$ source term
\begin{equation}
\begin{split}
    & \frac{b_{n}( r, \theta )}{r^{2}} =
        \Delta^{\ast}_{n} \psi_{0} -
        \dot{p}_{0} R_{n} - \dot{Y}_{0} \widetilde{R}_{n} \\
         & \;\;\; - \!\! \sum_{m = 1}^{n - 1} \binom{n}{m}
         \biggl(
            \dot{p}_{m} R_{n - m} + \dot{Y}_{m} \widetilde{R}_{n - m} -
                \Delta^{\ast}_{m} \psi_{n - m} \biggr),
\end{split}
\label{eq:nth.source}
\end{equation}
where the $m$-th order differential operator is defined as
\begin{equation}
    \Delta_{m}^{\ast} \equiv - \frac{1}{r} \frac{\partial}{\partial r}
        \biggl( r \widetilde{R}_{m} \frac{\partial}{\partial r} \biggr)
            - \frac{1}{r^{2}} \frac{\partial}{\partial \theta}
              \biggl( \widetilde{R}_{m} \frac{\partial}{\partial \theta}
              \biggr),
\label{eq:mth.GS.operator}
\end{equation}
while $\dot{p}_{m} = (d/d\varepsilon)^{m} \, \dot{p}(\psi)$, $\dot{Y}_{m} =
(d/d\varepsilon)^{m} \, \dot{Y}(\psi)$, $R_{m} =
(\partial/\partial\varepsilon)^{m} \, R$, and $\widetilde{R}_{m} =
(\partial/\partial\varepsilon)^{m} \, R^{-1}$, evaluated at $\varepsilon = 0$,
are radial functions~\cite{rodrigues.2004,rodrigues.2009}. Because
$\psi(r,\theta)$ is intended near the magnetic axis, the latter is made to
coincide with the origin demanding that $\nabla \psi$ vanishes at $r = 0$,
whence the conditions $\psi_{0}'(0) = \psi_{nk}(0) = \psi_{nk}'(0) = 0$.
Letting $\psi_{0}(0) = 0$ makes the poloidal flux vanish at the origin also,
being thus defined up to an additive constant.

To obtain $\psi_{0}(r)$, the source term in Eq.~\eqref{eq:0th.ode} is taken from
the poloidal-plane projection of the force balance, which yields the toroidal
current density
\begin{equation}
    - J_{(\phi)}(R,\psi) = R \dot{p}(\psi) + R^{-1} \dot{Y}(\psi)
\label{eq:toroidal.J}
\end{equation}
normalized to $\normpsi \big/ (\mu_{0} a^{2} R_{0})$. Its zeroth-order
term near the origin can be represented by the series
\begin{equation}
    - \Bigl[ \dot{p}_{0}(r) + \dot{Y}_{0}(r) \Bigr] =
                    \onaxisJ + \frac{J_{2}}{2} r^{2} + \cdots,
\end{equation}
where $\onaxisJ$ is the toroidal current density flowing on axis, $J_{1} = -
\bigl( d/dr \bigr) \bigl( \dot{p}_{0} + \dot{Y}_{0} \bigr) \big|_{0}$ vanishes
since $d/dr = \psi_{0}'(r) \, d/d\psi$, and $J_{2} = - \bigl( d^{2} \big/ dr^{2}
\bigr) \bigl( \dot{p}_{0} + \dot{Y}_{0} \bigr) \big|_{0}$, whence
\begin{equation}
    \psi_{0}(r) = \frac{\onaxisJ}{4} r^{2} + \frac{J_{2}}{32} r^{4} + \cdots.
\label{eq:psi0}
\end{equation}
Looking for the next solution $\psi_{1}(r,\theta)$, the source term
\begin{equation}
    b_{1}(r,\theta) = - r^{3} \biggl[ \dot{Y}_{0} - \dot{p}_{0} +
        \frac{1}{r^{2}} \frac{d}{dr} \Big( r^{2} \psi_{0}' \Bigr) \biggr]
        \cos\theta
\label{eq:source.b1}
\end{equation}
taken from~\eqref{eq:nth.source} shows that $\even{b}{10}$ and $\odd{b}{11}$
vanish, whereas
\begin{equation}
    \even{b}{11} = \tfrac{1}{2}
        \bigl[ \onaxisJ - 4 \dot{p}_{0}(0) \bigr] r^{3} + \cdots.
\end{equation}
Therefore, from Eq.~\eqref{eq:nkth.ode}, one finds $\even{\psi}{10} =
\odd{\psi}{11} = 0$ and
\begin{equation}
    \even{\psi}{11}= \tfrac{1}{16}
        \bigl[ \onaxisJ - 4 \dot{p}_{0}(0) \bigr] r^{3} + \cdots.
\label{eq:psi11}
\end{equation}
Similarly, $\even{b}{21}$, $\odd{b}{21}$, and $\odd{b}{22}$ are seen to
vanish also, while
\begin{equation}
    \even{b}{20}= B_{20} r^{4} + \cdots
    \quad \text{and} \quad
    \even{b}{22} = B_{22} r^{4} + \cdots.
\label{}
\end{equation}
As we shall see, the particular values of $B_{20}$ and $B_{22}$ are not
important in what follows. Again, Eq.~\eqref{eq:nkth.ode} yields
\begin{subequations}
\begin{align}
    \even{\psi}{20}(r) &= \frac{B_{20}}{16} r^{4} + \cdots, \\
    \even{\psi}{22}(r) &= \frac{\innC}{2} r^{2} + \frac{
        2 B_{22} - \sigma(0) \innC}{24} r^{4} + \cdots, \\
    \odd{\psi}{22}(r) &= \frac{\innS}{2} r^{2}
        - \frac{\sigma(0) \innS}{24} r^{4} + \cdots,
\end{align}
\label{eq:psi22}%
\end{subequations}
and $\even{\psi}{21}(r) = \odd{\psi}{21}(r) = 0$, where $\innC$ and $\innS$ are
integration constants whose meaning is discussed below.

Thus far, the solutions in Eqs.~\eqref{eq:psi0}, \eqref{eq:psi11}, and
\eqref{eq:psi22}, plus their combination in Eqs.~\eqref{eq:ansatz}, are valid
for any $r$ and $\varepsilon$, if all series converge and sufficient terms are
kept. The procedure outlined previously can thereby proceed to arbitrary powers
in $\varepsilon$. However, to get analytically tractable expressions, one must
truncate the power series in Eqs.~\eqref{eq:ansatz} somewhere. This is done
dropping all terms $\varepsilon^{\mu} r^{\nu}$ (where $\mu, \nu \geqslant 0$ are
integers) smaller than those leading in the combination $\varepsilon^{2}
\psi_{2}(r,\theta)$: First, let $\varepsilon \ll 1$ and $r \ll 1$; Then we find
$\varepsilon^{\mu+3} r^{\nu+2} \ll \varepsilon^{2} r^{2}$ and all
$\mathcal{O}(\varepsilon^{3})$ terms in Eqs.~\eqref{eq:ansatz}, which have the
form $\varepsilon^{\mu+3} r^{\nu+2}$, may be dropped when compared to
$\varepsilon^{2} r^{2} \bigl( \innC \cos 2\theta + \innS \sin 2\theta \bigr)
\big/ 4$; Further, the terms $\mathcal{O}(r^{4})$ in Eq.~\eqref{eq:psi22} are
also discarded since $\varepsilon^{2} r^{3+\nu} \ll \varepsilon^{2} r^{2}$; If,
in addition, one takes $r/\varepsilon \ll 1$, then $\varepsilon r^{3+\nu} \ll
\varepsilon^{2} r^{2}$ and the combination $\varepsilon \psi_{1}(r,\theta)$ may
be dropped altogether because its nonvanishing terms [listed in
Eq.~\eqref{eq:psi11}] are all of the form $\varepsilon r^{3+\nu}$; Likewise,
$r^{4+\nu} \ll \varepsilon^{2} r^{2}$ enables all terms $\mathcal{O}(r^{4})$ in
Eq.~\eqref{eq:psi0} to be discarded. Finally, the GS solution to lowest order in
aspect ratio ($\varepsilon \ll 1$) near the axis ($r \ll \varepsilon$), can be
written as
\begin{equation}
    \psi(r,\theta) \simeq \frac{\onaxisJ}{4} r^{2}
        + \frac{\varepsilon^{2}}{4} \Bigl(
            \innC r^{2} \cos 2\theta + \innS r^{2} \sin 2 \theta \Bigr).
\label{eq:GSsolution}
\end{equation}
Notice that $\varepsilon^{2} \innC$ and $\varepsilon^{2} \innS$ can be of the
same size as $\onaxisJ$.

\shortsection{Boundary conditions and plasma profiles}%
To better understand the physical meaning of the constants
\begin{equation}
    \innC \equiv \even{\psi}{22}{}''(0)
        \quad \text{and} \quad
            \innS \equiv \odd{\psi}{22}{}''(0)
\label{eq:innerCS}
\end{equation}
in Eq.~\eqref{eq:psi22}, let us write each $\psi_{nk}(r)$
as~\cite{rodrigues.2004,rodrigues.2005}
\begin{equation}
    \frac{\psi_{nk}(r)}{g_{k}(r)}  =
        \frac{\psi_{nk}(\rbound)}{g_{k}(\rbound)} -
            \int^{\rbound}_{r} \!\!\!\! \frac{du}{u g_{k}^{2}(u)}
                \int_{0}^{u} \frac{dv}{v} g_{k}(v) b_{nk}(v),
\label{eq:closed.integral}
\end{equation}
where $g_{k}(r)$ are homogeneous solutions of Eq.~\eqref{eq:nkth.ode} and
$\psi_{nk}(\rbound)$ are boundary conditions at $\rbound$. Define also
\begin{equation}
    \varsigma = \frac{g_{2}''(0)}{g_{2}(\rbound)}
    \quad \text{and} \quad
    \xi = \! \int^{\rbound}_{0} \!
        \frac{du}{u} \frac{g_{2}''(0)}{g_{2}^{2}(u)}
            \int_{0}^{u} \! \frac{dv}{v} g_{2}(v) \even{b}{22}(v),
\label{eq:CS.coefficients}
\end{equation}
and recall that $\odd{b}{22}$ vanishes whereas $\even{b}{22}$ does not. Hence,
\begin{equation}
    \innC = \varsigma \, \even{\psi}{22}(\rbound) - \xi
            \quad \text{and} \quad \innS = \varsigma \, \odd{\psi}{22}(\rbound).
\label{eq:CS.connection}
\end{equation}
Next, let $\psi( \rbound, \theta)$ be the angular distribution of the vacuum
poloidal-field flux, inferred from external magnetic
measurements~\cite{lee.1981,sartori.2003} along some radius $\rbound$ beyond the
plasma edge. Then, $\even{\psi}{22}(\rbound)$ and $\odd{\psi}{22}(\rbound)$ are
the leading terms of the complex $2^{\text{nd}}$-order Fourier coefficient
\begin{equation}
\begin{aligned}
    \outC + i \outS & \equiv
        \frac{1}{\pi} \int_{0}^{2 \pi} \!\! \!\!
            \psi( \rbound, \theta )e^{i 2 \theta} d\theta \\
        & = \frac{\varepsilon^{2}}{2} \Bigl[
            \even{\psi}{22}( \rbound ) +
            i \odd{\psi}{22}( \rbound ) \Bigr] +
        \mathcal{O}( \varepsilon^{3} )
\end{aligned}
\label{eq:outerCS}
\end{equation}
once the \textit{ansatz}~\eqref{eq:ansatz} has been recollected. Thus, $\innC$
and $\innS$ are proportional to $\outC$ and $\outS$, while $\varsigma$ and $\xi$
in Eq.~\eqref{eq:CS.connection} depend only on the plasma profiles $p(\psi)$ and
$Y(\psi)$.

\shortsection{On-axis up-down asymmetry}%
Setting $x = r \cos \theta$ and $y = r \sin \theta$, the equilibrium in
Eq.~\eqref{eq:GSsolution} becomes
\begin{equation}
    \psi(x,y) =
        \frac{\onaxisJ + \varepsilon^{2} \innC}{4} x^{2}
        + \frac{\varepsilon^{2} \innS}{2} x y
        + \frac{ \onaxisJ - \varepsilon^{2} \innC}{4} y^{2}.
\label{eq:quadratic.form}
\end{equation}
Its contours are ellipses, all with elongation $\kappa$ (the
major to minor-axis ratio) and tilted by some angle $\varphi$ such that
\begin{equation}
    \biggl| \frac{\kappa^{2} - 1}{\kappa^{2} + 1} \biggr| =
        \biggl| \frac{\varepsilon^{2}}{\onaxisJ} \biggr|
            \sqrt{\innC^{2} + \innS^{2}}
    \quad \text{and} \quad
    \tan 2 \varphi = \frac{\innS}{\innC}.
\label{eq:ellipses}
\end{equation}

Let a closed curve be defined as $F(x,y) = 0$, for some function
$F(x,y)$, along which $\nabla F$ does not vanish and
$\partial F \big/ \partial y = 0$ at two of its points only. These define the
limits $x_{1}$ and $x_{2}$ of the curve's projection on the line $y = 0$ and
such curve can thus be split into a top and bottom branch, respectively
$y_{\text{t}}(x)$ and $y_{\text{b}}(x)$, with $y_{\text{b}}(x) <
y_{\text{t}}(x)$ for $x_{1} < x < x_{2}$. So, we define the curve's
asymmetry as
\begin{equation}
    \eta = \!\! \int_{x_{1}}^{x_{2}} \tfrac{1}{2}
        \Bigl| y_{\text{b}}(x) + y_{\text{t}}(x) \Bigr| dx
        \Bigg/ \!\!
        \int_{x_{1}}^{x_{2}} \Bigl[ y_{\text{t}}(x) - y_{\text{b}}(x) \Bigr] dx.
\label{eq:eta}
\end{equation}
Once the top and bottom branches are sorted out from the quadratic
form~\eqref{eq:quadratic.form}, Eq.~\eqref{eq:eta} yields the single value
\begin{equation}
    \eta_{0} = \frac{\varepsilon^{2} \bigl| \innS \bigr|}{
        \pi \sqrt{ \onaxisJ^{2} -
            \varepsilon^{4} \bigl(\innC^{2} + \innS^{2} \bigr)}}
\label{eq:eta.onaxis}
\end{equation}
for all ellipses sufficiently near the axis. Additionally, if $\varepsilon^{4}
\bigl(\innC^{2} + \innS^{2} \bigr) \ll \onaxisJ^{2}$, the exact
Eq.~\eqref{eq:eta.onaxis} simplifies to
\begin{equation}
    \eta_{0}\approx \frac{\varepsilon^{2}}{\pi}
            \biggl| \frac{\innS}{\onaxisJ} \biggr|
        = \frac{2}{\pi} \biggl| \frac{ \varsigma \outS }{ \onaxisJ } \biggr|.
\label{eq:eta.onaxis.approx}
\end{equation}
Moreover, the relations in Eq.~\eqref{eq:ellipses} suffice to write $\eta_{0}$
as
\begin{equation}
    \eta_{0} = \frac{\eta_{0}^\text{max}}{\sqrt{1 + \cot^{2} 2 \varphi}},
    \quad \text{with} \quad
    \eta_{0}^\text{max} = \biggl| \frac{\kappa^{2} - 1}{4 \pi \kappa} \biggr|
\label{eq:maximum.eta}
\end{equation}
the maximum up-down asymmetry at constant $\kappa$, which is attained when
$\varphi =\pi/4$ and therefore $\innC \ll \innS$.

Since $J_{0}$ is a property of the core, whereas $\innC$ and $\innS$ depend
on the external field, they may be regarded as independent parameters
defining the equilibrium shape: Raising $\onaxisJ$ in Eq.~\eqref{eq:ellipses}
produces increasingly circular magnetic surfaces ($\kappa \sim 1$) and thus
suppresses locally the shaping imposed externally by $\outC$ and $\outS$ via
$\innC$ and $\innS$ in Eqs.~\eqref{eq:CS.connection} and~\eqref{eq:outerCS};
Conversely, the angle $\varphi$ is set by the external shaping only and does not
change with $\onaxisJ$.

The on-axis safety factor in the cylindrical approximation evaluates to $q_{0} =
2 a^{2} \normpsi^{-1} B_{0} \big/ \onaxisJ$. Therefore,
\begin{equation}
    \eta_{0} \approx \varsigma q_{0}
        \Biggl| \frac{\overline{\outS}}{\pi a^{2} B_{0}} \Biggr|
\label{eq:eta.onaxis.q}
\end{equation}
relates $\eta_{0}$ with the ratio of the poloidal-flux's odd Fourier harmonic
$\overline{\outS} = \normpsi \outS$ (in Wb) to the flux of the on-axis toroidal
field $B_{0}$ through the poloidal section. Of course,
Eqs.~\eqref{eq:eta.onaxis}, \eqref{eq:eta.onaxis.approx}, and
\eqref{eq:eta.onaxis.q} are all estimates of the particular
definition~\eqref{eq:eta}: higher $\eta$ does not necessarily imply more
momentum flux.

\begin{figure*}
\begin{center}
\includegraphics[scale=1.0]{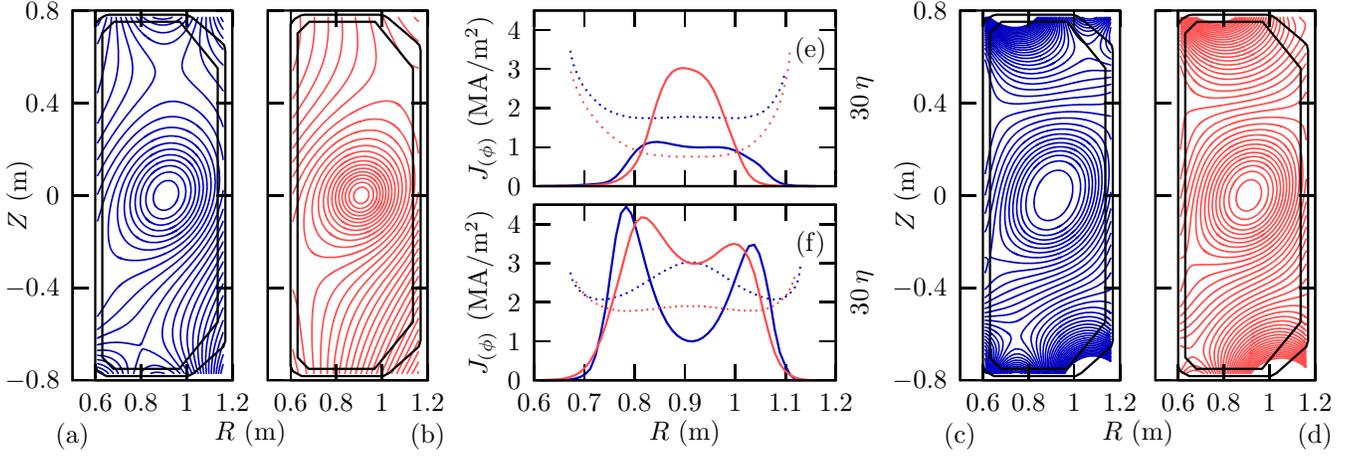}
\end{center}
\caption{\label{fig:equilibria}
    Magnetic equilibria for small [(a) and (b), left] and large [(c) and (d),
    right] $\plasmaI$: Equidistant contours start at the same value and are
    drawn in dark [(a) and (c)] and light tones [(b) and (d)] for low and high
    $\onaxisJ$ respectively; Midplane profiles of $J_{(\phi)}$ (solid
    lines) and $\eta$ (dotted lines) for small [(e), top] and large [(f),
    bottom] $\plasmaI$ scenarios follow the same tone rule.}
\end{figure*}

\shortsection{Analytic Solovev equilibrium}
The example bellow confirms the validity of the
\textit{ansatz}~\eqref{eq:ansatz} and the approximations leading to
Eq.~\eqref{eq:GSsolution}; it also illustrates the concepts embodied in
Eqs.~\eqref{eq:CS.connection}, \eqref{eq:CS.coefficients}, and
\eqref{eq:outerCS}. Let us consider the analytic Solovev
equilibrium~\cite{solovev.1968,cerfon.2010}
\begin{equation}
\begin{aligned}
    \psi(R,Z) &=
        - \tfrac{1}{2} U R^{2} \bigl( Z^{2} + \tfrac{1}{4} R^{2} \bigr) -
        \tfrac{1}{2} V \bigl( R^{2} \log R + Z^{2} \bigr) \\
        & \quad
        + a_{0} + a_{1} R^{2}
        + a_{2} Z + a_{3} Z R^{2} \\
        & \quad
        + a_{4} R^{2} \bigl( R^{2} - 4 Z^{2} \bigr)
        + a_{5} \bigl( Z^{2} - R^{2} \log R \bigr) \\
        & \quad 
        + a_{6} \Bigl[ 2 Z^{4} - 9 Z^{2} R^{2}
            + 3 R^{2} \bigl( R^{2} - 4 Z^{2} \bigr) \log R \Bigr],
\end{aligned}
\label{eq:solovev.RZ}
\end{equation}
which solves Eq.~\eqref{eq:grad.shafranov} for constant
$\dot{p}(\psi) = \varepsilon^{2} U$ and $\dot{Y}(\psi) = \varepsilon^{2} V$,
with $- \varepsilon^{-2} \Delta^{\ast} = \partial_{R} R^{-1} \partial_{R} +
R^{-1} \partial^{2}_{ZZ}$. Here, each arbitrary $a_{n}$ multiplies an
homogeneous GS solution. Of these coefficients, $a_{0}$, $a_{1}$, and $a_{2}$
are taken from the conditions $\psi(1,0) = \nabla \psi(1,0) = 0$, while $a_{3} =
- \tfrac{1}{4} \innS$,
\begin{equation}
\begin{aligned}
    a_{4} &=  2^{-10} \Bigl[
        23 \bigl( V - \innC \bigr) - 9 U - 27 c_{1} - 15 c_{2} \Bigr], \\
    a_{5} &= 2^{-7} \Bigl[
        7 V - 9 \bigl( U + c_{1} \bigr) - 39 \innC - 3 c_{2} \Bigr], \quad
    \text{and} \\
    a_{6} &= 2^{-8} \bigl( \innC - U - V + c_{1} + c_{2} \bigr)
\end{aligned}
\label{eq:coefficient.transform}
\end{equation}
are written in terms of four new arbitrary constants: $\innS$, $\innC$, $c_{1}$,
and $c_{2}$. The transformation~\eqref{eq:RZ.transformation}, after
setting $\varepsilon^{2} U = \dot{p}_{0}$ and $\varepsilon^{2} \bigl(U + V
\bigr) = - \onaxisJ$, turns Eq.~\eqref{eq:solovev.RZ} into
\begin{equation}
\begin{aligned}
    \psi & (r,\theta) = \frac{\onaxisJ}{4} r^{2}
        - \varepsilon
            \frac{\onaxisJ - 4 \dot{p}_{0}}{16} r^{3} \cos \theta
        + \frac{\varepsilon^{2}}{2} \Biggl[
            \frac{B_{20}}{16} r^{4} \\
        &
            + \biggl( \frac{\innC}{2} r^{2} + \frac{B_{22}}{12}
            r^{4} \biggr) \cos 2 \theta
            + \frac{\innS}{2} r^{2} \sin 2 \theta
        \Biggr] + \mathcal{O}(\varepsilon^{3}),
\end{aligned}
\label{eq:solovev.rtheta}
\end{equation}
where $B_{20} = - \tfrac{1}{4} \bigl( \onaxisJ + 8 \dot{p}_{0} \bigr)$ and
$B_{22} = - \tfrac{3}{8} \bigl( \onaxisJ + 4 \dot{p}_{0} \bigr)$. As expected,
its harmonics match the leading terms in the general Eqs.~\eqref{eq:psi0},
\eqref{eq:psi11}, and \eqref{eq:psi22}, since $\dot{p}_{0}$ and $\dot{Y}_{0}$
are here constant [and thus $\sigma(r) = 0$]. One finds also $g_{2}(r) \propto
r^{2}$, whence $\varsigma = 2 \rbound^{-2}$ and $\xi = \tfrac{1}{6} B_{22}
\rbound^{2}$ from Eqs.~\eqref{eq:CS.coefficients}.  The exact
solution~\eqref{eq:solovev.rtheta}, evaluated at $\rbound$, clearly illustrates
how $\innC$ and $\innS$ relate with $\outC$ and $\outS$ via the boundary
conditions, $\even{\psi}{22}(\rbound)$ and $\odd{\psi}{22}(\rbound)$, and
Eqs.~\eqref{eq:CS.connection} and \eqref{eq:outerCS}.

\shortsection{Numerical tokamak equilibria}
GS solutions are next computed for parameters typical of the Tokamak \`{a}
Configuration Variable (TCV)~\cite{hofmann.1994}, where evidence of momentum
flux induced by up-down asymmetry has been
reported~\cite{camenen.2010,camenen.2010a}. Convergent equilibria for finite
$\varepsilon$ and $r$ are obtained by dropping $\mathcal{O}(\varepsilon^{9})$
terms in the \textit{ansatz}~\eqref{eq:ansatz}, with Eqs.~\eqref{eq:nkth.ode}
and~\eqref{eq:0th.ode} yielding a set of equations for $n,k \leqslant 8$ which
are solved using the plasma models
\begin{subequations}
\begin{gather}
    p(\psi) = \tfrac{1}{2} P_{0} \Bigl[
        1 - \tanh \bigl( P_{1} \psi - P_{2} \bigr) \Bigl]
            \quad \text{and} \\
    \dot{p}(\psi) + \dot{Y}(\psi) = - \bigl( \onaxisJ + \alpha \psi \bigr)
        \exp \Bigl[ - \bigl( \gamma \psi \bigr)^{2} \Bigr],
\label{eq:plasma.models}
\end{gather}
\end{subequations}
where $P_{0}$, $P_{1} = 4.2$, $P_{2} = 2.65$, $\onaxisJ$, $\alpha$, and $\gamma$
are constants.

For a given plasma current $\plasmaI$, the normalized on-axis pressure
$P_{0}$ is chosen to keep the normalized $\beta$~\cite{wesson.1997}
\begin{equation}
    \beta_{\text{N}} =
        a B_{0} \frac{\beta}{\plasmaI} \sim
            \frac{2 \normpsi^{2}}{a B_{0} R_{0}^{2}}
                \frac{P_{0}}{\plasmaI} = 0.02 \:\text{Tm} \big/ \text{MA}
\label{eq:beta.condition}
\end{equation}
fixed, with $a = 0.24 \, \text{m}$, $R_{0} = 0.91 \, \text{m}$, and $B_{0} = 1.4
\, \text{T}$. The remaining parameters are chosen to fit the plasma inside the
vessel ($\alpha$, $\gamma$, and $\normpsi$) and to adjust its shape
[$\even{\psi}{22} \big|_{\text{B}}$, $\odd{\psi}{22} \big|_{\text{B}}$, and
$\even{\psi}{44} \big|_{\text{B}}$ at $\rbound = 1$, all other
$\psi_{nk}|_{\text{B}} = 0$, with $\psi_{nk}|_{\text{B}} \equiv
\psi_{nk}(r_{\text{B}})$]. These are listed in Table~\ref{tab:parameters} for
two scenarios (small and large $\plasmaI$), each with two configurations (low
and high $\onaxisJ$). The corresponding numerical equilibria are plotted in
Fig.~\ref{fig:equilibria}.
\begin{table}
\caption{\label{tab:parameters}
    Plasma parameters for numerical TCV-like equilibria: $\plasmaI$ (kA),
    $\normpsi$ (Wb), $P_{0}$ (kPa), and $\onaxisJ$ (MA/m$^{2}$); other
    parameters are dimensionless.}
\begin{ruledtabular}
\begin{tabular}{lrddrddccr}
    & \multicolumn{1}{c}{$\plasmaI$}
    & \multicolumn{1}{c}{$\normpsi$}
    & \multicolumn{1}{c}{$P_{0}$}
    & \multicolumn{1}{c}{$\onaxisJ$}
    & \multicolumn{1}{c}{$\alpha$}
    & \multicolumn{1}{c}{$\gamma$}
    & \multicolumn{1}{r}{$\even{\psi}{22} \big|_{\text{B}}$}
    & \multicolumn{1}{r}{$\odd{\psi}{22} \big|_{\text{B}}$}
    & \multicolumn{1}{r}{$\even{\psi}{44} \big|_{\text{B}}$} \\
    \hline
    (a) &  76 & 0.01  &  3.74 & 1 &  5.93 & 1.80 & 4 & 5 & 10 \\
    (b) &  76 & 0.01  &  3.74 & 3 & -7.25 & 1.40 & 3 & 5 & 10 \\
    (c) & 355 & 0.03 & 17.51 & 1 & 32.54 & 2.08 & 6 & 6 & 200 \\
    (d) & 355 & 0.03 & 17.51 & 3 & 10.80 & 1.44 & 6 & 6 & 200
\end{tabular}
\end{ruledtabular}
\end{table}

The profiles in Figs.~\ref{fig:equilibria} (e) and (f) agree with
Eq.~\eqref{eq:eta.onaxis}: Similar values of $\eta$ at the edge result in core
asymmetry which is lower for higher values of $\onaxisJ$. This asymmetry
suppression may be thought of as a competition between the imposed external
field, which affects $\innC$ and $\innS$, and the symmetric field induced
locally by $\onaxisJ$. Therefore, to increase the asymmetry on axis, one must
decrease $\onaxisJ$ without making $\plasmaI$ dwindle to undesirably low values.
Indeed, for any current-density profile with a global maximum on axis, the
constraint $\bigl| \onaxisJ \bigr| \gtrsim \bigl| \plasmaI \bigr| \big/ \bigl(
\pi a^{2} \bigr)$ places a limit on how much $\onaxisJ$ can be reduced for a
given $\plasmaI$. This problem is avoided using hollow current-density profiles,
as in Fig.~\ref{fig:equilibria}~(f), where most of the current flows off axis.
Besides affording lower $\onaxisJ$ values (among other advantages for
confinement and stability~\cite{kessel.1994}), hollow current profiles induce an
asymmetry build-up towards the core: the value of $\eta$ on axis grows to a
local maximum and may become larger than the one at the edge.

\begin{table}
\caption{\label{tab:onaxis.asymmetry}
On-axis asymmetry: estimated values ($\eta_{0}$) against numerical ones
[$\eta(0)$]. All variables are dimensionless.}
\begin{ruledtabular}
\begin{tabular}{lddddd}
    & \multicolumn{1}{c}{$\onaxisJ$}
    & \multicolumn{1}{c}{$\varsigma$}
    & \multicolumn{1}{c}{$\outS$}
    & \multicolumn{1}{c}{$\eta_{0}$}
    & \multicolumn{1}{c}{$\eta(0)$} \\
    \hline
    (a) & 6.58 & 3.41 & 0.174 & 0.0574 & 0.0593 \\
    (b) & 19.7 & 4.44 & 0.174 & 0.0249 & 0.0252 \\
    (c) & 2.07 & 1.50 & 0.209 & 0.0962 & 0.101  \\
    (d) & 6.21 & 2.87 & 0.209 & 0.0613 & 0.0636
\end{tabular}
\end{ruledtabular}
\end{table}

The estimate in Eq.~\eqref{eq:eta.onaxis.approx} is tested in
Table~\ref{tab:onaxis.asymmetry}, where the values of $\onaxisJ$, $\varsigma$,
and $\outS$ [with $\varsigma$ and $\outS$ computed from
Eqs.~\eqref{eq:CS.coefficients} and~\eqref{eq:outerCS}] for each configuration
in Table~\ref{tab:parameters} are used to evaluate $\eta_{0}$. The latter is
found to agree rather well with the numerical value $\eta(0)$, which is
computed directly from Eq.~\eqref{eq:eta} for magnetic surfaces near the axis.

\begin{figure}
\begin{center}
\includegraphics[scale=1.0]{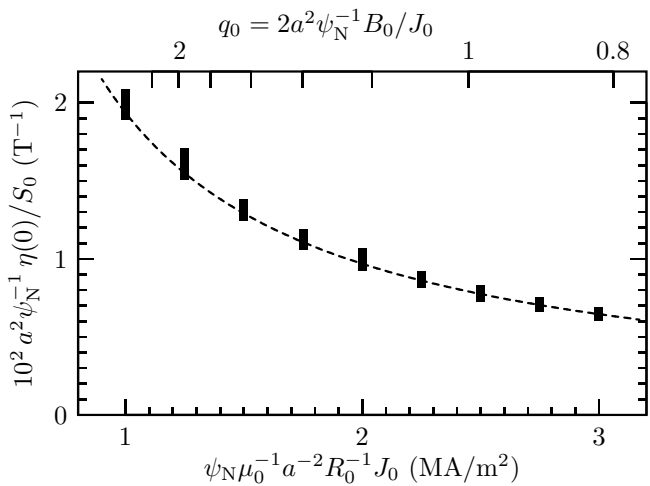}
\end{center}
\caption{\label{fig:asymmetry}
    Ratio of computed $\eta(0)$ to $\innS$ (squares) and analytic curve $\eta(0)
    \big/ \innS \approx \varepsilon^{2} \big/ \bigl( \pi \onaxisJ \bigr)$
    (dashed line).}
\end{figure}

Lastly, a large number ($\sim270$) of equilibria are computed for a broader set
of parameters, with the on-axis current density ranging between $1$ and $3 \,
\text{MA/m}^{2}$, while $\plasmaI = 75$, $200$, and $350$~kA. Other parameters
are varied within reasonable bounds to keep the plasma inside the vessel. For
each equilibrium, $\innS$ is evaluated as $\odd{\psi}{22}{}''(0)$ and the ratio
$\eta(0) \big/ \innS$ is plotted against $\onaxisJ$ in Fig.~\ref{fig:asymmetry}.
The agreement with Eq.~\eqref{eq:eta.onaxis.approx} is manifest. Also, the
dispersion of computed results around the curve, due to nonzero $\varepsilon^{4}
\bigl( \innC^{2} + \innS^{2} \bigr)$ in Eq.~\eqref{eq:eta.onaxis}, is higher for
smaller $\onaxisJ$.

\shortsection{Conclusions}%
An analytic up-down asymmetric equilibrium was developed near the magnetic axis,
which depends on three parameters only: $\onaxisJ$, $\innC$, and $\innS$. The
asymmetry was quantified and found to be established by the ratio of $\innS
\propto \varsigma \outS$ (which measures the ability of the external-field's odd
perturbation $\outS$ to propagate into the core) to the on-axis current density
$\onaxisJ$. The results were tested with a Solovev equilibrium and with
numerical equilibria computed for TCV parameters.  An important finding is that
hollow current profiles (and thus reverse magnetic shear) are seen to improve
on-axis asymmetry values, which can even be larger than those at the edge. These
findings may enhance the intrinsic rotation in the core, thus improving plasma
confinement.

\begin{acknowledgments}
This work was supported by EURATOM and carried out within the framework of the
European Fusion Development Agreement. IST activities were also supported by
``Funda\c{c}\~{a}o para a Ci\^{e}ncia e Tecnologia'' through project
Pest-OE/SADG/LA0010/2011. The views and opinions expressed herein do not
necessarily reflect those of the European Commission. J.~Ball and F.~I.~Parra
were supported by US DOE Grant No.~DE-SC008435.
\end{acknowledgments}

\end{document}